\documentclass[lettersize,journal]{IEEEtran}
\usepackage{amsmath,amsfonts}
\usepackage{algorithmic}
\usepackage{algorithm}
\usepackage{array}
\usepackage[caption=false,font=normalsize,labelfont=sf,textfont=sf]{subfig}
\usepackage{textcomp}
\usepackage{stfloats}
\usepackage{url}
\usepackage{verbatim}
\usepackage{booktabs}
\usepackage{graphicx}
\usepackage{listings}
\usepackage{cite} % Required for better citation handling
\usepackage{parskip}
\usepackage{amsthm}
\usepackage{xcolor}
\usepackage{tcolorbox}
\usepackage{multirow}
\newtheorem{definition}{Definition}
\begin{document}

\title{Sample-Efficient LLM-Based Detection of Malicious Web Server Logs with Forensically Explainable Reasoning}

% \author{IEEE Publication Technology,~\IEEEmembership{Staff,~IEEE,}}

\author{Bernhard Kneip,
        Nhien-An Le-Khac
        and Hong-Hanh Nguyen-Le}

\maketitle

\begin{abstract}
Forensic analysis of web server logs demands both accurate detection and human-readable explanations that can satisfy legal requirements. We present CEF-Log, a context-enhanced few-shot chain-of-thought prompting strategy for Large Language Models that addresses this dual requirement. CEF-Log embeds expert investigative methodology through a structured five-step reasoning template, enabling the model to learn \textit{how} to analyze logs rather than \textit{what} patterns to memorize. Experimental evaluation demonstrates that CEF-Log achieves an F1-score of 0.99 on the CSIC 2010 dataset using only four examples while providing a $10\times$ improvement in sample efficiency compared to other prompting-based methods. We also introduce ForenWebLog, a new dataset that incorporates real-world attacks and multi-step attack sequences for comprehensive evaluation. Qualitative analysis confirms that CEF-Log generates traceable, accurate explanations suitable for forensic documentation, addressing the critical "black-box" limitation of traditional machine learning approaches.
\end{abstract}

\section{Introduction} \label{sec:introduction}
Web server logs constitute essential evidence in cybercrime investigations, capturing detailed traces of HTTP requests including client IP addresses, requested resources, and response codes \cite{davidoff2012network, qi2017cloud, chen2019empirical}. When organizations discover security breaches, often days after the initial attack, forensic investigators must rapidly sift through these logs to identify evidence linking malicious actions to perpetrators \cite{Kent_Chevalier_Grance_Dang_2006, zhong2018learning}. This triage process is critically time-sensitive: data protection regulations impose strict limits on the retention of IP-related data \cite{europa}, and cross-border investigations introduce procedural delays. If leads are not pursued promptly, evidence may be irretrievably lost.

The scale of modern web server logs, often millions of entries daily, renders manual analysis impractical \cite{xu2009detecting, qi2017cloud, chen2019empirical}. Traditional automation approaches rely on forensic experts using custom scripts and predefined rule sets, such as YARA \cite{virustotalYARAPattern} and Sigma \cite{sigmahqExploreSigma}, to identify known attack patterns. However, these methods are inherently limited by the rules applied; without prior knowledge of the specific attack vector, identifying the correct ruleset remains a significant challenge. Machine Learning (ML) and Deep Learning (DL) techniques such as Decision Trees, SVMs, and Hidden Markov Models have shown promise in anomaly detection \cite{saleem2020web, cao2017machine, he2016experience, nedelkoski2020self}, but present significant limitations in forensic contexts. These supervised methods require substantial labeled training data, which is scarce given the sensitive nature of attack logs and continuous emergence of new patterns \cite{landauer2023deep}. Additionally, the opacity of ML models, commonly known as the "black box" problem, poses fundamental challenges where reasoning must be transparent, reproducible, and legally admissible \cite{le2022log}.

Large Language Models (LLMs) \cite{chang2024survey} present a compelling alternative for forensic log analysis. LLMs can process unstructured textual data while generating human-readable explanations, which is valuable when investigators must articulate reasoning to legal stakeholders \cite{scanlon2023chatgpt}. Moreover, LLMs demonstrate few-shot learning capabilities, potentially enabling effective analysis without extensive labeled datasets \cite{brown2020language}. Recent studies have explored LLM applications in digital forensics, including memory forensics triage \cite{oh2024volgpt}, system log anomaly detection \cite{guan2024logllm, liu2024logprompt, zhang2025llm}, and report generation \cite{michelet2024chatgpt}, through fine-tuning-based and prompt engineering-based strategies. 

% \textcolor{red}{Highlight focus on unstable logs.}

However, these methods exhibit fundamental limitations when applied to web server log analysis. First, existing LLM-based methods target operational system logs (e.g., HDFS \cite{xu2009online}, BGL \cite{oliner2007supercomputers}, Thunderbird \cite{oliner2007supercomputers}) rather than web server logs, addressing fundamentally different detection challenges. System logs record internal runtime events with structured templates, where anomalies manifest as deviations from expected execution sequences \cite{du2017deeplog}. In contrast, web server logs capture external HTTP requests containing adversarial payloads that are deliberately crafted to evade detection, including SQL injection, cross-site scripting, and path traversal attacks, often employing sophisticated obfuscation techniques. Second, fine-tuning-based methods require tens of thousands of labeled training samples, conflicting with forensic investigations where labeled attack examples are scarce during initial triage. Third, prompt engineering-based methods suffer from high false positive rates due to limited domain-specific knowledge, failing to provide structured, forensically sound reasoning. Our experimental results in Table \ref{tab:prompting_limitations} summarize the limitations of standard prompting techniques when evaluating on CSIC 2010 \cite{torrano2009self}. 

% Despite their potential, the naive application of standard prompting paradigms to web server log triage reveals significant limitations. Zero-shot prompting frequently suffers from low recall, failing to identify a significant portion of actual attacks, while standard few-shot prompting requires a volume of labeled examples that is often impractical to obtain during the initial stages of a real-world investigation. Furthermore, these basic strategies often fail to account for the multi-step nature of modern attacks, processing log entries in isolation rather than as a coherent sequence of events. Our experimental results in Table \ref{tab:prompting_limitations} summarize the quantitative and qualitative limitations observed when evaluating on CSIC 2010 \cite{torrano2009self}. 

\begin{table*}[ht]
    \centering
    \caption{Limitations of Standard Prompting Paradigms in Forensic WebLog Triage}
    \label{tab:prompting_limitations}
    \resizebox{\textwidth}{!}{%
    \begin{tabular}{l|c|c|c|p{7cm}}
    \toprule
    \textbf{Prompting Paradigm} & \textbf{Data Requirement} & \textbf{Average Recall} & \textbf{Average F1-Score} & \textbf{Primary Forensic Limitation} \\
    \midrule
    Zero-Shot & None & 0.59 & 0.73 & High rate of false negatives; lacks contextual depth for complex attacks \\
    \midrule
    Few-Shot (Standard) & High (40 examples) & 0.96 & 0.98 & Requires excessive pre-labeled data, which is rarely available in active investigations \\
    \midrule
    Role-Based & None & 0.52 & 0.68 & Marginal improvement over zero-shot; fails to provide significant performance gains \\
    \midrule
    \textbf{Ours} & \textbf{Low (4 examples)} & \textbf{0.95} & \textbf{0.99} & - \\
    \bottomrule
    \end{tabular}%
    }
\end{table*}

To address these challenges, we propose \textbf{CEF-Log}, a context-enhanced few-shot chain-of-thought prompting strategy, which systematically integrates expert forensic knowledge into the reasoning process. Our approach constructs domain-specific reasoning templates guiding the LLM through structured analysis: log parsing, forensic feature extraction, contextual interpretation, and classification with justification. Each few-shot example includes detailed reasoning traces demonstrating expert analysis, from identifying suspicious URL patterns to recognizing obfuscation techniques. This effectively transfers domain expertise to the LLM, enabling expert-level analysis with minimal examples. Our experiments demonstrate that CEF-CoT achieves performance comparable to 40-example few-shot prompting using only 4 examples, representing a $10 \times$ improvement in sample efficiency. Furthermore, to demonstrate the effectiveness of our proposed method, we also collect a new realistic dataset derived from current web server logs, synthetically enriched with multi-step attack patterns that better represent modern threat landscapes.

\textbf{Contributions.} Our contributions include:
\begin{itemize}
    \item A novel CEF-Log prompting method achieving F1-score of 0.99 on the CSIC 2010 dataset with only four examples, outperforming traditional ML and existing prompting strategies;
    \item A new realistic dataset, namely ForenWebLog, derived from current web server logs, synthetically enriched with multi-step attack patterns such as web shell execution and sophisticated Denial of Service (DoS) attempts. We also validate CEF-Log on this dataset to show its effetiveness against real-world attacks;
    \item A structured evaluation of LLM reasoning, confirming that these models can generate detailed, human-readable justifications that overcome the "black-box" limitations inherent in earlier automated systems;
    \item Comprehensive comparison against ML baselines (SVM, Random Forest) and prompting techniques (zero-shot, few-shot, CoT) on two datasets using accuracy, precision, recall, and F1-score metrics.
\end{itemize}

\textbf{Paper Organization.} The remainder of this paper is organized as follows. Section \ref{sec:background} provides background on web server logs, forensic triage, and the anomaly detection problem formulation. Section \ref{sec:related-work} reviews related work in ML/DL-based web attack detection and LLM-based log analysis. Section \ref{sec:methodology} details the CEF-Log methodology, including the reasoning template design, few-shot example construction, and the ForenWebLog dataset collection framework. Section \ref{sec:experiments} presents experimental results and analysis. Section \ref{sec:conclusion} concludes the paper.

\section{Background}
\label{sec:background}

This section provides foundational concepts essential for understanding the challenges of malicious web server log detection. We first distinguish web server logs from system logs, then introduce the forensic triage process, and finally formalize the anomaly detection problem with particular attention to the distinction between stable and unstable log distributions.

\subsection{Web Server Logs}
\label{subsec:webserverlogs}

Web server logs are generated by HTTP servers (e.g., Apache, Nginx, IIS) that capture details of client requests and server responses. A typical log entry in Common Log Format (CLF) or Combined Log Format contains the following fields: client IP address, timestamp of the request, request line, response code (HTTP status code returned by the server), response size, referrer (the URL from which the request originated), and User-Agent (client software identification string) \cite{suneetha2009identifying}. An example log entry is shown in Fig.~\ref{fig:log_example}.

\begin{figure}[ht]
\centering
\begin{lstlisting}[
  basicstyle=\small\ttfamily,
  frame=single,
  breaklines=true,
  columns=fullflexible
]
192.168.1.100 - - [15/Dec/2024:10:15:32 +0000]
"GET /search?q=admin'%20OR%201=1-- HTTP/1.1"
200 5234 "http://example.com" "Mozilla/5.0"
\end{lstlisting}
\caption{Example web server log entry in Combined Log Format containing a SQL injection attempt in the query parameter.}
\label{fig:log_example}
\end{figure}

Web server logs differ fundamentally from system logs (e.g., HDFS~\cite{xu2009detecting}, BGL, Thunderbird~\cite{oliner2007supercomputers}) in several critical aspects, as summarized in Table~\ref{tab:log_comparison}. System logs record internal runtime events generated by software components, typically following predefined templates where dynamic values (e.g., process IDs, memory addresses) are inserted into static text patterns \cite{hadadi2025llm}. Anomalies in system logs manifest as deviations from expected execution sequences, which are an unusual ordering of events or the presence of unexpected log templates.

\begin{table}[ht]
\centering
\caption{Comparison Between System Logs and Web Server Logs}
\label{tab:log_comparison}
\begin{tabular}{|p{1.8cm}|p{2.8cm}|p{2.8cm}|}
\hline
\textbf{Characteristic} & \textbf{System Logs} & \textbf{Web Server Logs} \\
\hline
Source & Internal application/OS events & External HTTP requests \\
\hline
Structure & Template-based with parameters & Semi-structured with free-form payloads \\
\hline
Content & Runtime states, errors, diagnostics & URLs, parameters, headers, user input \\
\hline
Threat Model & Operational anomalies, system failures & Adversarial attacks, deliberate evasion \\
\hline
Anomaly Type & Sequence deviations, unexpected events & Malicious payloads, injection attacks \\
\hline
\end{tabular}
\end{table}

In contrast, web server logs capture external requests that may contain adversarial payloads deliberately crafted to evade detection. Attack payloads such as SQL injection (\texttt{' OR 1=1--}), cross-site scripting (\texttt{<script>alert(1)</script>}), and path traversal (\texttt{../../etc/passwd}) are intentionally obfuscated using encoding techniques (URL encoding, Unicode normalization, case manipulation) to bypass security controls. This adversarial nature necessitates detection approaches that understand attack semantics rather than merely pattern matching against known templates.

% \subsection{Early Forensic Triage}
% \label{subsec:forensictriage}

% Digital forensic investigations follow a structured methodology comprising identification, preservation, collection, examination, analysis, and reporting phases \cite{Kent_Chevalier_Grance_Dang_2006}. \textit{Forensic triage} refers to the initial rapid assessment phase where investigators prioritize evidence sources and identify leads that warrant deeper investigation \cite{zhong2018learning}.

% In the context of web server security incidents, triage serves several critical functions: (1) \textit{rapid threat identification} which means quickly distinguishing attack traffic from normal operations to determine incident scope; (2) \textit{timeline reconstruction} which means establishing the sequence of malicious actions from initial access to data exfiltration; and (3) \textit{evidence prioritization} which means identifying log entries that provide forensic value for attribution and legal proceedings.

\subsection{Anomaly Detection on Web Server Logs}
\label{subsec:anomalydetection}

\begin{figure}[ht]
\centering
    \includegraphics[width=\columnwidth]{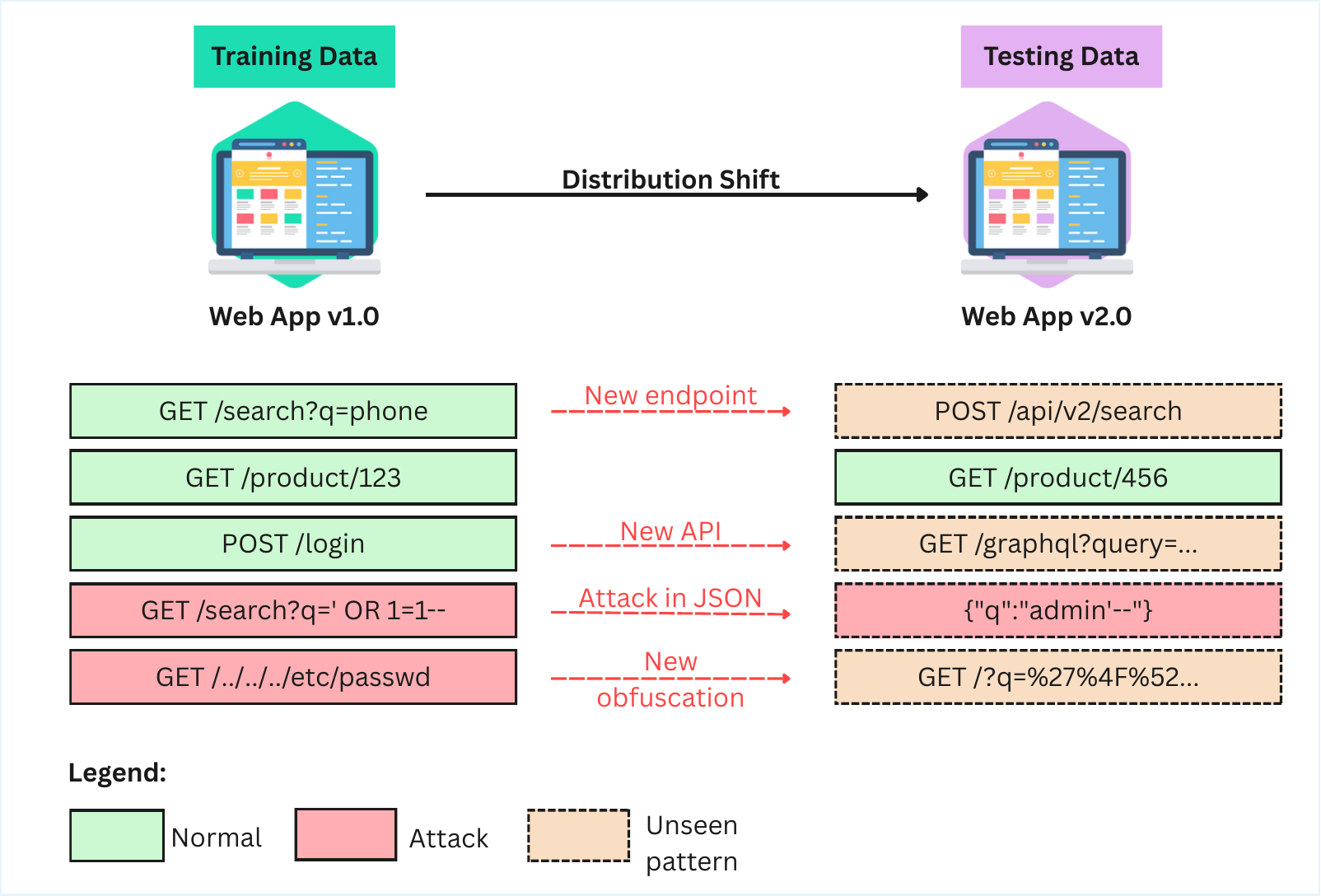}
    \caption{Illustration of unstable web server logs. Training data from Web App v1.0 follows distribution $\mathcal{P}_{train}$, while testing data from v2.0 follows a different distribution $\mathcal{P}_{test}$ due to application updates, evolving attack techniques, and new obfuscation methods. Dashed boxes indicate patterns not present in training data.}
    \label{fig:unstable_logs}
\end{figure}

Anomaly detection in web server logs constitutes a binary classification task: given a log entry or sequence of entries, determine whether it represents normal traffic or malicious activity. We formalize the distinction between \emph{stable} and \emph{unstable} log distributions, adapting concepts from recent work in system log analysis \cite{hadadi2025llm} to the web server domain.

\begin{definition}[Stable Logs]
\label{def:stable}
Web server logs are \textit{stable} when both training and testing data are drawn from the same underlying distribution $\mathcal{P}$. The distribution encompasses the set of legitimate endpoints and their usage patterns, normal parameter value distributions, typical user-agent distributions and access patterns, and known attack types and their manifestations.
\end{definition}

In stable conditions, a detection model trained on historical data encounters test instances that reflect the same web application structure, user behavior, and threat landscape.

\begin{definition}[Unstable Logs]
\label{def:unstable}
Web server logs are \textit{unstable} when training $\mathcal{P}_{train}$ and testing $\mathcal{P}_{test}$ data are drawn from different underlying distributions ($\mathcal{P}_{train} \neq \mathcal{P}_{test}$). 
\end{definition}

% The distribution shifts may occur due to:
% \begin{itemize}
%     \item \textbf{Application evolution}: Web applications undergo continuous development, such as new endpoints are added, existing APIs are modified or deprecated, and parameter schemas change. A model trained on logs from application version 1.0 may encounter entirely new URL patterns in version 2.0.
%     \item \textbf{Attack evolution}: Adversaries continuously adapt their techniques to evade detection. New vulnerability disclosures spawn novel attack variants; obfuscation techniques evolve to bypass existing signatures; and sophisticated attackers chain multiple techniques in ways not represented in training data.
%     \item \textbf{Environmental changes}: Infrastructure modifications (new load balancers, CDNs, or API gateways), geographic expansion, and shifts in user demographics alter traffic distributions.
%     \item \textbf{Temporal drift}: Even without explicit changes, the distribution of web traffic exhibits natural temporal variation due to seasonal patterns, marketing campaigns, or external events.
% \end{itemize}

\begin{definition}[Anomaly Detection on Stable Logs]
\label{def:slad}
Let $D^{train}$ denote the training dataset and $D^{test}_{S}$ denote the testing dataset, both sampled from the same distribution $\mathcal{P}$. The stable anomaly detection task aims to learn a classifier $f: \mathcal{X} \rightarrow \{0, 1\}$ from $D^{train}$ that accurately predicts labels on $D^{test}_{S}$.
\end{definition}

\begin{definition}[Anomaly Detection on Unstable Logs]
\label{def:ulad}
Let $D^{train}$ be sampled from distribution $\mathcal{P}_{train}$ and $D^{test}_{U}$ be sampled from a different distribution $\mathcal{P}_{test}$ where $\mathcal{P}_{train} \neq \mathcal{P}_{test}$. The unstable anomaly detection task aims to learn a classifier $f$ from $D^{train}$ that generalizes to $D^{test}_{U}$ despite the distribution shift.
\label{def:anomaly_unstable}
\end{definition}

The unstable log detection problem is substantially more challenging because models must generalize beyond their training distribution. Traditional ML approaches, which learn decision boundaries optimized for the training distribution, often fail catastrophically when distributions shift. This motivates our approach of leveraging LLMs with domain-specific prompting, which can reason about attack semantics rather than relying solely on statistical patterns that may not transfer across distributions. Fig. \ref{fig:unstable_logs} illustrates unstable web server logs. 

\section{Related Work} \label{sec:related-work}

\subsection{Machine Learning (ML) and Deep Learning (DL) in Web-based Attack Detection}
Researchers have utilized ML techniques for anomaly detection in log data. Early statistical approaches employed textual feature extraction methods to model attack patterns. Ren et al. \cite{ren2018web} pioneered the use of Bag-of-Words representations for web payload analysis, subsequently applying Hidden Markov Models (HMMs) to capture sequential dependencies in normal traffic behavior. Building on this idea, Cao et al. \cite{cao2017machine} introduced a hierarchical classification framework that combines a decision tree with HMM-based behavioral modeling. Their two-stage architecture first discriminates between benign and suspicious entries using structural features before employing trained HMMs to identify deviations from legitimate traffic distributions. Complementary work by Saleem \cite{saleem2020web} demonstrated the efficacy of term frequency-inverse document frequency (TF-IDF) vectorization for multiclass attack classification. 

The emergence of DL has enabled more sophisticated representation learning for web attack detection. Tekerek \cite{tekerek2021novel} proposed a payload-based anomaly detection system that identifies anomalous patterns in HTTP traffic. By translating HTTP payloads into binary pixel maps, this method leverages the powerful pattern-recognition capabilities of CNNs to detect unknown threats that signature-based systems might miss. Eunaicy et al. \cite{eunaicy2022web} explored the use of the Long Short-Term Memory model for capturing temporal dependencies in web attack sequences. Pan et al. \cite{pan2019detecting} designed an autonomic intrusion detection system using end-to-end DL with stacked denoising autoencoders to recognize anomalies without manual feature engineering. By leveraging the Robust Software Modeling Tool, the system calculates reconstruction errors to establish a threshold between normal execution and malicious activities such as SQL injection and cross-site scripting.

Despite these advances, empirical evaluations \cite{inuwa2024comparative, elmrabit2020evaluation} consistently reveal a fundamental limitation: both traditional ML classifiers (e.g., SVMs, k-NN) and DL architectures require substantial quantities of labeled training data to achieve acceptable detection rates. This requirement poses significant challenges in forensic contexts where ground-truth labels are scarce, and attack patterns continuously evolve.

\subsection{Large Language Models (LLMs) for Anomaly System Log Detection}
Recent studies have explored LLM applications in log analysis through two primary paradigms: fine-tuning-based and prompting-based approaches. 

\textbf{Fine-tuning-based methods} adapt pre-trained LLMs to specific log datasets through supervised learning. LogLLM \cite{guan2024logllm} employs a dual-encoder architecture combining BERT for semantic vector extraction with Llama for sequence classification. This method requires a three-step training procedure across tens of thousands of samples. LLMeLog \cite{he2024llmelog} utilizes ChatGPT 3.5 to enrich log events with domain knowledge and anomaly tendency through a structured prompt designed for in-context learning. These enriched events are then used to fine-tune a pre-trained BERT model via a hierarchical metric loss, which subsequently generates event representations for training a transformer-based anomaly detection model. LLM-LADE \cite{zhang2025llm} fine-tunes LLaMA3-8B using Low-Rank Adaptation (LoRA) with GPT-4-augmented training data. To maintain performance in dynamic environments, LLM-LADE incorporates focal loss to address class imbalance and an online knowledge base for continuous optimization. Moreover, LLM-LADE also explains anomalies by outputting label-cause pairs that associate a detection label with a natural language description of the anomaly's underlying cause. FLEXLOG \cite{hadadi2025llm} is a hybrid approach that integrates a fine-tuned Mistral with traditional ML models using ensemble learning. To handle unstable system logs effectively, the method incorporates Retrieval-Augmented Generation to provide contextual information and a cache mechanism to enhance inference efficiency. FLEXLOG effectively uses only 500 fine-tuning samples, but can achieve up to a 13 percentage point increase in F1 score. 

\textbf{Prompting-based} methods leverage the zero-shot or few-shot capabilities of LLMs through carefully designed prompts. LogGPT \cite{qi2023loggpt} utilizes a prompt construction strategy consisting of a task description, format statement, human knowledge injection, and input sequences to guide ChatGPT in identifying anomalies and providing explanations. However, this approach encounters significant scalability bottlenecks; as log grouping windows expand, the context window limitations of the LLM often lead to information loss. RAGLog \cite{pan2024raglog} addresses these constraints by incorporating a Retrieval-Augmented Generation framework. Specifically, RAGLog dynamically retrieves semantically similar historical logs and known anomaly patterns from a vector database to provide the model with "in-context" grounding without over-leveraging the prompt's token limit. Building on this, LogPrompt \cite{liu2024logprompt} employs Chain-of-Thought (CoT) prompting to improve anomaly detection by guiding the model through reasoning steps. While avoiding the training data requirements of fine-tuning approaches, these methods remain prone to high false-positive rates. This is primarily due to the LLMs' lack of domain-specific knowledge and their tendency to hallucinate under ambiguous log syntax \cite{huang2025survey}. 

Our work fills the current gaps of existing prompting-based approaches by integrating structured forensic domain knowledge through context-enhanced reasoning templates. Our work also achieves sample efficiency comparable to zero-shot methods while maintaining detection performance approaching fine-tuning approaches, requiring only 4 annotated examples compared to the tens of thousands needed by LogLLM or LLM-LADE. 

\section{Methodology} \label{sec:methodology}
\subsection{Problem Formulation}
We formalize the task of malicious web server log detection as a binary classification problem augmented with explainability requirements essential for forensic investigations. 

\textbf{Input Representation.} Let $\mathcal{L}= \{l_1, l_2, \dots, l_n\}$ denote a sequence of web server log entries. Each log entry $l_i$ is represented as a tuple: 
\begin{equation}
    l_i = (ip_i, t_i, m_i, u_i, s_i, r_i, a_i, H_i),
\end{equation}
where $ip_i, t_i, m_i \in \{\text{GET}, \text{POST}, \text{PUT}, \text{DELETE}, \cdots\}, u_i, \\ s_i, r_i, a_i, H_i = \{(k_j, v_j) \}^{|H_i|}_{j=1}$ denotes the client IP address, the timestamp, the HTTP status code, the referrer URL, the User-Agent string, and additional HTTP headers as key-value pairs (e.g., cookies, content-type, custom headers), respectively. 

\textbf{Output Specification.} Unlike conventional anomaly detection systems that produce only binary labels, forensic triage demands interpretable outputs. We define the output for each log entry $l_i$ as a pair: 
\begin{equation}
    o_i = (y_i, e_i), 
\end{equation}
where $y_i \in \{\text{Benign}, \text{Attack}\}$ is the classification label, and $e_i$ is a natural language explanation from LLM that explains the reasoning behind the classification. The explanation $e_i$ must satisfy the following forensic requirements:
\begin{enumerate}
    \item \textbf{Traceability}: $e_i$ must reference specific elements within $l_i$ that influenced the classification decision.
    \item \textbf{Accuracy}: $e_i$ must correctly identify attack vectors (e.g., SQL injection, XSS, path traversal) when present.
\end{enumerate}

\textbf{Classification Objective.} Given a log sequence $\mathcal{L}$ and a small set of labeled examples $\mathcal{D}^{few} = \{(l_j, y_j, e_j)\}^k_{j=1}$ where $k \ll |\mathcal{L}|$, the objective is to learn a classification function:
\begin{equation}
    f: \mathcal{L} \times \mathcal{D}^{few} \rightarrow \{(\hat{y}_i, \hat{e}_i)\}^n_{i=1}
\end{equation}
In our approach, $k=4$, representing a significant reduction compared to traditional few-shot methods requiring $k \geq 40$ examples.

Following Definition \ref{def:anomaly_unstable}, web server log analysis inherently operates under unstable conditions. The training examples $\mathcal{D}^{few}$ are drawn from distribution $\mathcal{P}_{train}$ representing known attack patterns, while the test logs $\mathcal{L}$ may follow a different distribution $\mathcal{P}_{test}$ due to: (1) Novel attack variants not represented in training examples; (2) Application-specific URL structures and parameter schemas; and (3) Evolving obfuscation techniques designed to evade known signatures.

Our approach addresses this distribution shift by encoding transferable forensic reasoning principles rather than memorizing specific attack signatures, enabling generalization to previously unseen attack patterns. 

\begin{figure*}[ht]
\centering
    \centering
    \includegraphics[width=\textwidth]{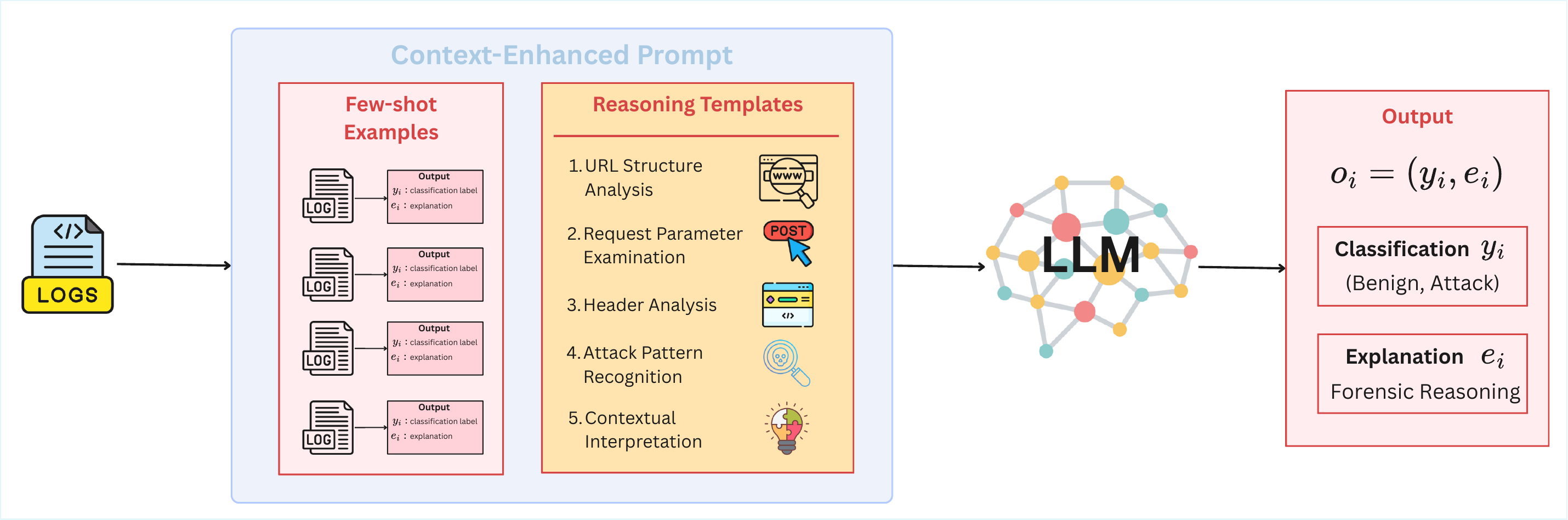}
    \caption{Overview of CEF-Log for forensic web log classification. Few-shot examples paired with a five-step reasoning template guide the LLM through structured analysis, producing classifications with forensically sound explanations.}
    \label{fig:overview}
\end{figure*}

\subsection{CEF-Log: Context-Enhanced Few-Shot Chain-of-Thought Prompting}
This section presents CEF-Log, our proposed prompting strategy that systematically integrates forensic domain expertise into LLM-based log analysis.

\subsubsection{Overview}
CEF-Log combines three complementary techniques: (1) chain-of-thought prompting to encourage structured reasoning, (2) few-shot learning to demonstrate expert analysis patterns, and (3) domain-specific context enhancement to inject forensic knowledge. Fig.~\ref{fig:overview} illustrates the overall architecture.

The prompt construction pipeline consists of four stages. First, we define a \textit{reasoning template} $\mathcal{T}$ that decomposes log analysis into discrete analytical steps reflecting expert investigative methodology. Second, we construct a \textit{few-shot example set} $\mathcal{D}^{few}$ containing representative attack instances paired with detailed reasoning traces. Third, we assemble the complete prompt by combining task instructions, the reasoning template, few-shot examples, and the target log entries. Finally, the LLM processes this prompt to generate classifications with accompanying explanations.

Formally, let $\mathcal{M}$ denote the LLM, $\mathcal{T}$ the reasoning template, $\mathcal{D}^{few} = \{(l_j, r_j, y_j)\}_{j=1}^{k}$ the few-shot examples where $r_j$ represents the reasoning trace for log entry $l_j$, and $\mathcal{L}_{test}$ the target log entries. The CEF-Log classification function is defined as:
\begin{equation}
f_{CEF}(\mathcal{L}_{test}) = \mathcal{M}(\mathcal{P}(\mathcal{T}, \mathcal{D}^{few}, \mathcal{L}_{test}))
\label{eq:cef_function}
\end{equation}
where $\mathcal{P}(\cdot)$ denotes the prompt assembly function that constructs the complete input prompt.

\subsubsection{Reasoning Template Design}
The reasoning template $\mathcal{T}$ structures the analytical process into five sequential steps. This designed template serves two purposes: it guides the LLM through systematic examination of all relevant features, and it produces explanations that align with forensic documentation standards. 

These five steps are derived from established forensic investigation practices, reflecting the cognitive workflow that experienced analysts follow when triaging web server logs. By encoding this domain expertise directly into the prompt structure, we achieve two objectives: guiding the LLM toward systematic and thorough analysis that minimizes oversight of subtle attack indicators, while simultaneously ensuring that the generated explanations align with forensic documentation standards required for legal proceedings. 

\textbf{Step 1: HTTP Method and URL Structure Analysis.} The first analytical step examines the request line components. The LLM evaluates whether the HTTP method is appropriate for the target resource, presence of suspicious patterns in the URL path (e.g., directory traversal sequences such as \texttt{../}, access attempts to sensitive files like \texttt{.htaccess} or \texttt{web.config}), anomalous URL encodings that may indicate obfuscation attempts, and query parameter structure and naming conventions. This step addresses static attacks that attempt to access restricted resources and identifies obvious structural anomalies.

\textbf{Step 2: Request Parameter Examination.} The second step performs a deep inspection of all request parameters, encompassing URL query strings, POST body data, and other payload-carrying fields. The analysis includes extraction and decoding of URL-encoded content, identification of injection patterns (i.e., SQL keywords, script tags, system commands), detection of parameter tampering (i.e., modified field names, unexpected data types), and recognition of fuzzing indicators (i.e., boundary values, malformed inputs). This step targets dynamic attacks that exploit input validation vulnerabilities.

\textbf{Step 3: HTTP Header Analysis.} The third step examines HTTP headers for anomalous content, including \textit{cookie values} for session manipulation or injection attempts, \textit{User-Agent strings} for known malicious tools or inconsistencies, \textit{referrer headers} for cross-site request patterns, and \textit{custom headers} that may carry attack payloads (e.g., Log4Shell attacks via \texttt{X-Forwarded-For}). 

\textbf{Step 4: Attack Pattern Recognition.} The fourth step synthesizes observations from previous steps to identify known and unknown attack categories:
\begin{itemize}
    \item \textit{SQL Injection}: presence of SQL syntax, logical operators, comment sequences
    \item \textit{Cross-Site Scripting (XSS)}: script tags, event handlers, JavaScript URIs
    \item \textit{Path Traversal}: directory navigation sequences, absolute path references
    \item \textit{Command Injection}: shell metacharacters, command chaining operators
    \item \textit{Authentication Attacks}: credential stuffing patterns, session fixation attempts
    \item \textit{Suspicious Attacks}: target unknown attacks, but LLM suspects it is an attack
\end{itemize}
The LLM leverages its pre-trained knowledge of attack signatures while remaining alert to obfuscated variants.

\textbf{Step 5: Contextual Interpretation.} It considers whether the observed patterns could result from legitimate user behavior and evaluates the plausibility of the request within normal application workflows. The analysis also examines correlation with other requests from the same source for multi-step attack detection. Finally, an overall risk assessment integrates all preceding observations. This step is critical for reducing false positives by distinguishing genuine attacks from benign anomalies.

\subsubsection{Few-shot Example Construction}
Unlike conventional few-shot learning \cite{qi2023loggpt, pan2024raglog}, where examples provide a representative sample for pattern memorization, CEF-Log's examples serve to demonstrate the application of the reasoning methodology. The primary objective is to teach the LLM how to systematically analyze logs rather than what specific patterns to detect. This fundamental shift, \emph{from pattern memorization to forensic analysis}, explains why CEF-Log requires only 4 examples compared to the 40+ typically needed for standard few-shot prompting.

\textbf{Example Structure.} Each few-shot example $(l_j, r_j, y_j) \in \mathcal{D}^{few}$ comprises three components:

\begin{itemize}
    \item \textit{Log Entry} $l_j$: The complete HTTP request in structured format (JSON representation preserving all fields)
    \item \textit{Reasoning Trace} $r_j$: Step-by-step analysis following template $\mathcal{T}$, explicitly documenting observations and inferences at each step
    \item \textit{Classification} $y_j$: Final label with summary justification
\end{itemize}

The reasoning trace $r_j$ is critical for effective knowledge transfer. Rather than merely providing input-output pairs, the detailed reasoning demonstrates \textit{how} to analyze logs, enabling the LLM to generalize the analytical methodology to novel instances. 
% Detailed examples are provided in \textcolor{red}{Appendix}. 

\textbf{Example Set Composition.} We identify four fundamental detection challenges that, when addressed through examples, enable generalization across the attack landscape: (1) \textit{payload decoding}, recognizing obfuscated content through URL encoding or other techniques; (2) \textit{structural anomaly detection}, identifying subtle parameter modifications indicative of automated tools; (3) \textit{semantic validation}, detecting inputs that violate domain constraints; and (4) \textit{injection pattern recognition}, identifying embedded executable content. By providing examples that represent these four challenges, we provide a sufficient demonstration of the reasoning methodology for the LLM to generalize effectively. Our experimental results in Section V validate this design: CEF-Log with $k=4$ examples achieves an F1-score of $0.98$, matching the performance of standard few-shot prompting with  $k=40$ examples while requiring $10 \times$ fewer labeled samples.

\subsubsection{Prompt Assembly}

The complete prompt is assembled by concatenating four components in sequence: task instructions, reasoning template with few-shot examples, analysis directive, and target log entries.

\textbf{Component 1: Task Instructions.} The prompt begins with a clear task specification:

\begin{tcolorbox}[
    colback=white,
    colframe=blue!70!black,
    boxrule=1pt,
    arc=2pt,
    left=6pt,
    right=6pt,
    top=6pt,
    bottom=6pt
]
\textit{``Analyze the following HTTP requests to determine if they are malicious. First, I will provide you with examples demonstrating step-by-step analysis methodology.''}
\end{tcolorbox}
This establishes the classification objective and signals that worked examples will follow.

\textbf{Component 2: Reasoning Template with Examples.} Each few-shot example is presented with its complete reasoning trace, structured according to template $\mathcal{T}$. The reasoning explicitly labels each analytical step:

\begin{tcolorbox}[
    colback=white,
    colframe=blue!70!black,
    boxrule=1pt,
    arc=2pt,
    left=6pt,
    right=6pt,
    top=6pt,
    bottom=6pt
]
\textit{``Step 1: URL Analysis --- The URL \texttt{http://...} contains...''}\\[2pt]
\textit{``Step 2: Analyze request data --- The method is POST, examining Post-Data reveals...''}\\[2pt]
\textit{``...''}\\[2pt]
\textit{``Final Classification: Attack''}
\end{tcolorbox}

This structured presentation reinforces the analytical methodology and provides concrete instantiations of each reasoning step.

\textbf{Component 3: Analysis Directive.} Following the examples, we include instructions for analyzing the target data:
\begin{tcolorbox}[
    colback=white,
    colframe=blue!70!black,
    boxrule=1pt,
    arc=2pt,
    left=6pt,
    right=6pt,
    top=6pt,
    bottom=6pt
]
\textit{``Now, analyze the following data using the same step-by-step reasoning process and provide a final classification. Also incorporate your own knowledge about additional attack patterns not covered in the examples.''}
\end{tcolorbox}
The final sentence is important: it explicitly authorizes the LLM to leverage its pre-trained knowledge of attack patterns beyond those demonstrated in the examples, enabling generalization to novel attack types.

\textbf{Component 4: Target Log Entries.} The log entries requiring classification are provided in structured JSON format, preserving all fields from the original log representation. Each entry includes a unique record identifier to enable unambiguous mapping between inputs and outputs.

\textbf{Output Format Specification.} To facilitate automated processing of results, we constrain the LLM output to structured JSON format containing, for each log entry: the record identifier, classification label, and concise reasoning summary. We enforce this format using the native structured output capabilities of the target LLM APIs (e.g., Pydantic schemas for Google Gemini, tool-based formatting for Anthropic Claude).

Algorithm~\ref{alg:cef_log} summarizes the complete CEF-Log classification procedure.

\begin{algorithm}[t]
\caption{CEF-Log Classification}
\label{alg:cef_log}
\begin{algorithmic}[1]
\REQUIRE Log entries $\mathcal{L}_{test} = \{l_1, \ldots, l_n\}$, Few-shot examples $\mathcal{D}^{few}$, Reasoning template $\mathcal{T}$, LLM $\mathcal{M}$
\ENSURE Classifications with explanations $\{(\hat{y}_i, \hat{e}_i)\}_{i=1}^{n}$
\STATE $prompt \leftarrow \textsc{TaskInstruction}()$
\FOR{each $(l_j, r_j, y_j) \in \mathcal{D}^{few}$}
    \STATE $prompt \leftarrow prompt \,\|\, \textsc{FormatExample}(l_j, r_j, y_j, \mathcal{T})$
\ENDFOR
\STATE $prompt \leftarrow prompt \,\|\, \textsc{AnalysisDirective}()$
\STATE $prompt \leftarrow prompt \,\|\, \textsc{FormatLogEntries}(\mathcal{L}_{test})$
\STATE $response \leftarrow \mathcal{M}(prompt)$
\STATE $\{(\hat{y}_i, \hat{e}_i)\}_{i=1}^{n} \leftarrow \textsc{ParseResponse}(response)$
\RETURN $\{(\hat{y}_i, \hat{e}_i)\}_{i=1}^{n}$
\end{algorithmic}
\end{algorithm}

% \subsubsection{Multi-Step Attack Detection}

% While the preceding components address single-request classification, sophisticated attacks often span multiple requests. CEF-Log extends to multi-step attack detection through contextual prompting that presents related log entries together, enabling the LLM to identify cross-request patterns.

% For multi-step detection, we modify the analysis directive to include:
% %
% \begin{quote}
% \textit{``Consider relationships between requests from the same IP address or targeting related resources. Multi-step attacks such as web shell uploads followed by execution, or brute force sequences, manifest as patterns across multiple entries.''}
% \end{quote}

% When processing log sequences, we preserve temporal ordering and source IP associations, allowing the LLM to recognize patterns such as:
% %
% \begin{itemize}
%     \item \textbf{Denial of Service}: Repeated identical requests from a single source within short time intervals
%     \item \textbf{Brute Force}: Multiple authentication attempts with varying credentials
%     \item \textbf{Web Shell Attacks}: File upload requests followed by execution requests to the uploaded resource
% \end{itemize}

% Our experimental evaluation demonstrates that CEF-Log successfully identifies these multi-step patterns, with the LLM explicitly referencing earlier requests in its reasoning (e.g., \textit{``This request attempts to execute a file created by the attack in Record ID 7''}).

\subsection{ForenWebLog Dataset Collection Framework}

\begin{figure}[ht]
\centering
    \includegraphics[width=\columnwidth]{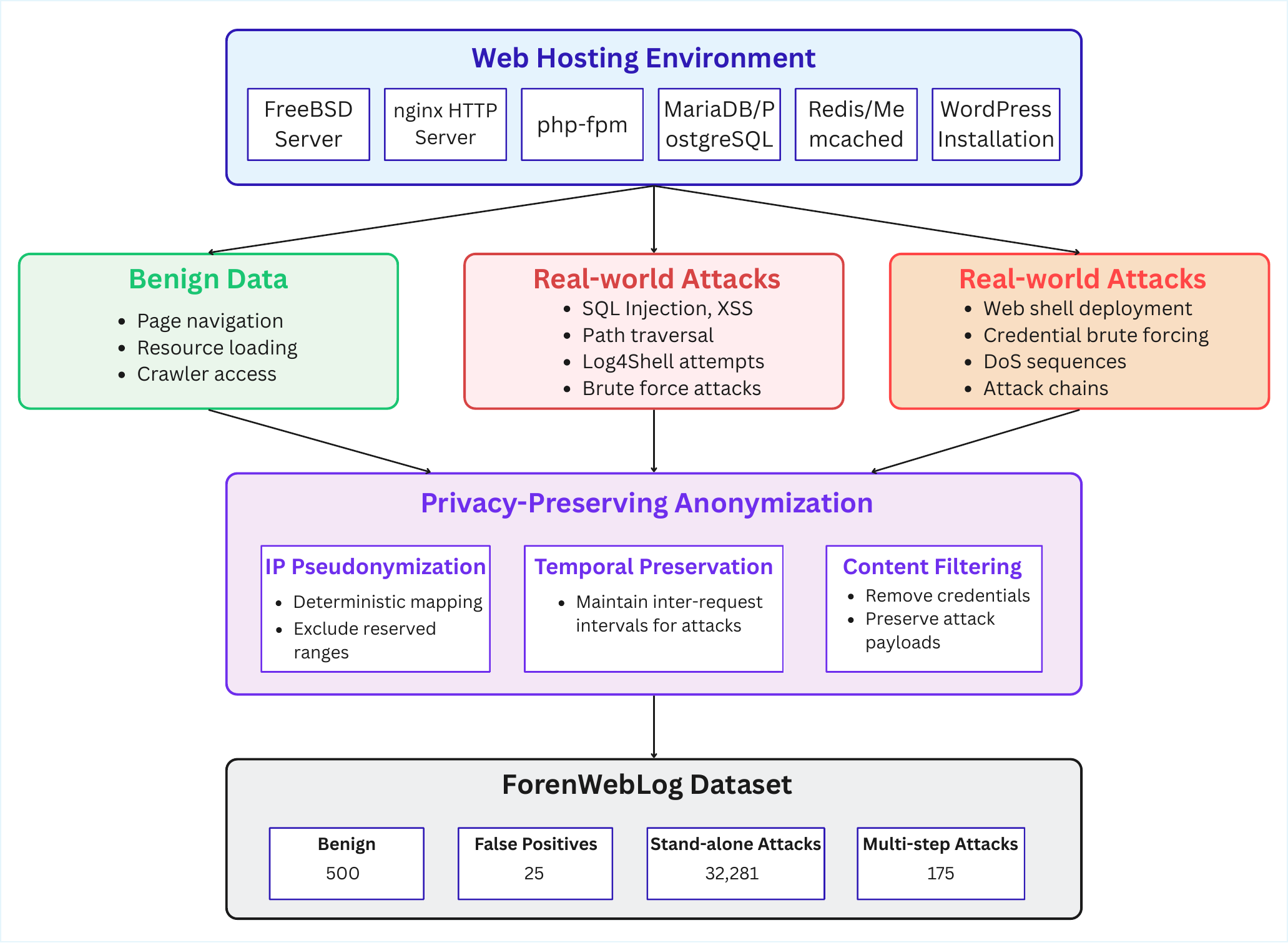}
    \caption{Illustration of ForenWebLog Dataset Collection Framework.}
    \label{fig:unstable_logs}
\end{figure}

Existing web server log datasets exhibit significant limitations for evaluating LLM-based forensic analysis. The CSIC 2010 dataset~\cite{torrano2009self}, while widely adopted, lacks temporal information essential for detecting time-based attacks such as DDoS and brute force attempts. Furthermore, each request stands in isolation without contextual relationships to subsequent entries, limiting the evaluation of multi-step attack detection capabilities. The ECML/PKDD 2007 dataset \cite{lirmmAttackChallenge} suffers from excessive randomization that renders legitimate traffic indistinguishable from attacks without ground-truth labels. To address these gaps, we constructed a new dataset derived from production web server logs, incorporating both authentic attack traffic and synthetically generated multi-step attack sequences.

\subsubsection{Collection Framework}

Our data collection leverages a production web hosting environment operational for several years. The server runs FreeBSD \cite{dinh2005freebsd} with \textit{nginx} as the HTTP server and \textit{php-fpm} for dynamic content, with MariaDB/PostgreSQL backends \cite{kenler2015mariadb, postgresql1996postgresql} and Redis/Memcached caching \cite{carlson2013redis}. The logging configuration generates separate log files for each hosted web application, stored in a centralized location with automated rotation managed by the operating system. Logs follow the Combined Log Format, capturing: client IP address, request timestamp with timezone, HTTP method and URI, response status code, response size in bytes, referrer URL, and User-Agent string. This standardized format ensures compatibility with common analysis tools while preserving all fields necessary for forensic investigation.

The primary data source is a WordPress installation serving the local volunteer fire department website---selected because it serves only informational content (preserving privacy) while attracting substantial attack traffic as a high-value target. Data extraction employed Unix utilities and custom Python scripts over a multi-month collection period.

\subsubsection{Data Sources}
The primary data source is a WordPress installation serving as the public website for a local volunteer fire department. There are two primary reasons for this choice: (1) it serves exclusively informational content with no user-specific URL patterns that could compromise privacy; (2) WordPress installations are high-value targets due to widespread deployment and frequently outdated configurations, attracting substantial attack traffic. 

Our dataset integrates three complementary data sources to ensure comprehensive coverage of both common and sophisticated attack patterns:

\textbf{Benign Data.} Legitimate requests including page navigation, resource loading, and crawler access from established services (Googlebot, GPTBot, AhrefsBot).

\textbf{Real-World Attack Traffic.} Authentic attacks harvested from production logs: SQL injection, cross-site scripting, path traversal, command injection, Log4Shell attempts, and brute force attacks with varying obfuscation levels.

\textbf{Synthetic Multi-Step Attacks.} While real-world logs contained various attack types, certain sophisticated multi-step attack patterns were underrepresented due to log rotation policies and the tendency of automated attack tools to abort upon initial failure. We therefore synthesized additional attack sequences based on documented exploitation techniques, including:
\begin{itemize}
    \item \textit{Web shell deployment}: Path traversal or file upload exploitation followed by remote code execution via the uploaded script
    \item \textit{Credential brute forcing}: Repeated authentication attempts against login endpoints with varying credentials
    \item \textit{Denial of Service}: High-frequency requests targeting resource-intensive endpoints (e.g., XML-RPC, search functionality)
\end{itemize}
These synthetic sequences preserve realistic timing characteristics and request structures derived from observed attack patterns.

\subsubsection{Privacy-preserving Anonymization}
Our anonymization pipeline implements: (1) \textit{IP pseudonymization}, (2) \textit{temporal preservation}, and (3) \textit{content filtering}.

\textbf{IP Address Pseudonymization.} All client IP addresses are replaced with randomly generated addresses using a deterministic mapping within each log file. This approach preserves the critical property that requests from the same source maintain consistent identifiers, enabling detection of multi-step attacks and behavioral patterns. Generated addresses explicitly exclude reserved ranges (private networks, multicast, loopback) to prevent false attribution of suspicious characteristics. 

\textbf{Temporal Preservation.} For single-request entries, timestamps are randomized independently. However, for multi-step attack sequences, we preserve relative timing by randomizing only the initial timestamp while maintaining inter-request intervals. This design choice is critical: temporal patterns often distinguish malicious automation (sub-second intervals between requests) from human behavior, and serve as essential features for detecting DDoS and brute force attacks.

\textbf{Content Filtering.} URL paths and query parameters are retained to preserve attack payload semantics. However, we manually reviewed entries to remove any inadvertently captured credentials or session tokens, replacing them with placeholder values while maintaining structural validity.

The anonymization process is formally defined as:
\begin{equation}
    l'_i = \mathcal{A}(l_i) = (\phi(\text{ip}_i), \tau(t_i, \mathcal{S}_i), m_i, u_i, s_i, r_i, a_i, H_i)
\end{equation}
where $\phi: \mathcal{IP} \rightarrow \mathcal{IP}'$ is the IP pseudonymization function with consistent mapping, $\tau$ adjusts timestamps based on sequence membership $\mathcal{S}_i$, and remaining fields are preserved unchanged.

\begin{table}[ht]
\centering
\caption{Multi-Step Attack Categories in Our Dataset}
\label{tab:multistep}
\begin{tabular}{lccl}
\toprule
\textbf{Category} & \textbf{Steps} & \textbf{Variants} & \textbf{Key Indicators} \\
\midrule
DoS & 50+ & 3 & Request frequency, payload size \\
Brute Force & 20--50 & 3 & Login endpoints, same source IP \\
Web Shell & 2 & 3 & Upload then execute pattern \\
\bottomrule
\end{tabular}
\end{table}

\subsubsection{Multi-Step Attack Enrichment}
We categorize multi-step attacks into three categories. A distinguishing feature of our dataset is the inclusion of multi-step attack sequences that require contextual analysis across multiple log entries. We define three categories of enriched attack patterns: 
\begin{itemize}
    \item \textbf{Denial of Service (DoS) Sequences.} (1) XML-RPC abuse targeting WordPress installations with repeated POST requests to \texttt{/xmlrpc.php}; (2) application-layer flooding via oversized POST payloads to dynamic endpoints; and (3) resource exhaustion through rapid requests to computationally expensive pages. Each sequence contains 50 attack requests interspersed with legitimate traffic to evaluate detection under realistic conditions.
    \item \textbf{Brute Force Sequences.} Authentication attacks are represented through repeated login attempts against WordPress (\texttt{wp-login.php}) and webmail interfaces (Roundcube).
    \item \textbf{Web Shell Attack Chains.} The most sophisticated category involves two-phase attacks: initial exploitation followed by execution of the deployed payload. Exploitation vectors include: (1) path traversal combined with PEAR command injection \cite{stasinopoulos2019commix} to write PHP files; (2) WebDAV PUT \cite{goland1999http} requests to upload executable content; and (3) exploitation of vulnerable file upload endpoints (e.g., Gravity Forms plugin).
\end{itemize}

\subsubsection{Dataset Composition and Statistics} 
The final dataset comprises 32,981 log entries organized into four categories as detailed in Table~\ref{tab:dataset_stats}.

\begin{table}[ht]
    \centering
    \footnotesize
    \caption{Dataset Composition}
    \label{tab:dataset_stats}
    \begin{tabular}{lrl}
        \toprule
        \textbf{Category} & \textbf{Count} & \textbf{Description} \\
        \midrule
        Benign & 500 & Legitimate traffic \\
        False Positives & 25 & HTTPS-on-HTTP attempts \\
        Stand-alone Attacks & 32,281 & Single-request attacks \\
        Multi-step Attacks & 175 & Attack sequences \\
        \midrule
        \textbf{Total} & \textbf{32,981} & \\
        \bottomrule
    \end{tabular}
\end{table}

\section{Experiments} \label{sec:experiments}

\subsection{Experiments Setup}

\subsubsection{Datasets} We evaluate CEF-Log on two datasets representing distinct forensic scenarios.

\textbf{CSIC 2010 Dataset}~\cite{torrano2009self}. This dataset contains over 36,000 HTTP requests targeting an e-commerce application, labeled as normal or anomalous. We use stratified sampling to extract balanced test sets of 100 entries (50 normal, 50 anomalous) across five random seeds, ensuring reproducibility while maintaining class distribution.

\textbf{ForenWebLog Dataset}. We also evaluate our proposed method against real-world attacks in our collected dataset - ForenWebLog. For each experimental seed, we sample: 50 benign requests, 40 stand-alone attacks, 2 multi-step attack sequences (comprising multiple log entries each), and 5 false positive candidates (HTTPS-on-HTTP connection attempts that appear suspicious but are benign). This composition specifically tests the model's ability to distinguish sophisticated attacks from benign anomalies.

\subsubsection{LLM Models} We conduct experiments with two popular LLMs via API. For \textbf{Google Gemini Flash 2.5}, we use it for all experiments. The model supports up to 1,048,576 input tokens and 8,192 output tokens, enabling analysis of substantial log batches. We also use \textbf{Claude 3.7 Sonnet} for cross-validation on selected experiments to confirm that improvements stem from our proposed prompting CEF-Log method. 

\subsubsection{Metrics} We adopt standard classification metrics derived from the confusion matrix:

\begin{equation}
\text{Accuracy (Acc.)} = \frac{TP + TN}{TP + TN + FP + FN}
\end{equation}

\begin{equation}
\text{Precision (Prec.)} = \frac{TP}{TP + FP}, \quad \text{Recall (Rec.)} = \frac{TP}{TP + FN}
\end{equation}

\begin{equation}
\text{F1-Score} = \frac{2 \cdot \text{Precision} \cdot \text{Recall}}{\text{Precision} + \text{Recall}}
\end{equation}

\noindent where $TP$, $TN$, $FP$, and $FN$ denote true positives, true negatives, false positives, and false negatives, respectively. In the forensic context, \textit{Recall} measures the proportion of actual attacks correctly identified (critical for ensuring no threats are missed), while \textit{Precision} measures the reliability of attack classifications (important for minimizing investigator workload on false alarms). All experiments are repeated across five random seeds, and we report mean and standard deviation to assess stability.

\subsubsection{Baselines} We compare CEF-Log against both prompting-based and machine learning baselines. 

\textbf{Prompting Baselines}. We compare with zero-shot, role-based zero-shot, and few-shot prompting baselines. For the few-shot method, we evaluate with different $k$ values ($k \in \{1, 5, 10, 15, 20, 25, 30, 35, 40\}$). We also compare our method with other prompting-based method for log data, including: LogPrompt \cite{guan2024logllm} and LogGPT \cite{qi2023loggpt}. 

\textbf{Machine Learning (ML) Baselines}. We compare with Support Vector Machine (SVM), Random Forest (RF), and Logistic Regression (LR). Note that, ML baselines are trained on 40 labeled examples per seed (20 benign, 20 attack) to enable fair comparison with few-shot prompting at equivalent data requirements.

\subsubsection{Prompts} Detailed prompts used for the zero-shot prompting, role-based prompting, and few-shot prompting are:
\begin{itemize}
    \item Zero-shot prompting: \textit{"Analyze the following HTTP requests and decide, if it is malicious or not. Use the words 'Benign' and 'Attack' for classification. Here is the data: \{experimental\_data\}"}
    \item Role-based prompting: \textit{You are a \{role\_name\}. Analyze the following HTTP requests and decide, if it is malicious or not. Use the words ’Benign’ and ’Attack’ for classification. Here is the data: \{experimental\_data\}}. We test with three main roles:  computer forensic expert ($\texttt{Forensic}$), cybercrime investigator ($\texttt{Cybercrime}$), and security specialist ($\texttt{Security}$). 
    \item Few-shot prompting: \textit{Analyze the following HTTP requests and decide, if it is malicious or not. Use the words 'Benign' and 'Attack' for classification. The following is a list of examples that you should refer to before giving the final decision: \{list\_of\_examples\}}. 
\end{itemize}

\subsubsection{Implementation Details} Our framework is implemented in Python using Google Generative AI SDK and Anthropic API. Structured JSON outputs are enforced via Pydantic schemas (Gemini) and tool-based formatting (Claude). Log entries are converted to JSON to preserve all HTTP fields, with custom parsers for each dataset format. Each condition is evaluated across five random seeds with stratified sampling; we report mean and standard deviation.

\subsection{Experimental Results}
\subsubsection{Performance Comparison with Standard Prompting Baselines}
\begin{table}[ht]
    \centering
    \caption{Performance Comparison of Prompting Strategies on CSIC 2010 Dataset (Gemini Flash 2.5)}
    \label{tab:standard_prompting_baselines}
    \begin{tabular}{lcccc}
    \hline
    \textbf{Method} & \textbf{Acc.} & \textbf{Prec.} & \textbf{Rec.} & \textbf{F1} \\
    \hline
    Zero-Shot & 0.79 & 0.99 & 0.59 & 0.73 \\
    Role-Based (Forensic) & 0.76 & 0.99 & 0.52 & 0.68 \\
    Role-Based (Cybercrime) & 0.76 & 1.00 & 0.53 & 0.68 \\
    Role-Based (Security) & 0.75 & 0.99 & 0.50 & 0.66 \\
    Few-Shot ($k$=10) & 0.88 & 0.99 & 0.76 & 0.86 \\
    Few-Shot ($k$=20) & 0.89 & 0.99 & 0.78 & 0.87 \\
    Few-Shot ($k$=40) & 0.98 & 0.99 & 0.96 & 0.98 \\
    \textbf{CEF-Log ($k$=4)} & \textbf{0.98} & \textbf{0.99} & \textbf{0.98} & \textbf{0.98} \\
    \hline
    \end{tabular}
\end{table}

\begin{table}[ht]
    \centering
    \caption{Cross-Validation: Gemini Flash 2.5 vs. Claude 3.7 Sonnet on CSIC 2010}
    \label{tab:cross_validation}
    \begin{tabular}{llcccc}
        \hline
        \textbf{Method} & \textbf{Model} & \textbf{Acc.} & \textbf{Prec.} & \textbf{Rec.} & \textbf{F1} \\
        \hline
        \multirow{2}{*}{Zero-Shot} & Gemini & 0.79 & 0.99 & 0.59 & 0.73 \\
         & Claude & 0.76 & 0.99 & 0.53 & 0.67 \\
        \hline
        \multirow{2}{*}{Role-Based} & Gemini & 0.76 & 0.99 & 0.52 & 0.68 \\
         & Claude & 0.77 & 0.99 & 0.54 & 0.68 \\
        \hline
        \multirow{2}{*}{CoT (Refined)} & Gemini & 0.95 & 0.94 & 0.97 & 0.95 \\
         & Claude & 0.84 & 0.98 & 0.69 & 0.78 \\
        \hline
        \multirow{2}{*}{\textbf{CEF-Log ($k$=4)}} & Gemini & \textbf{0.98} & \textbf{0.99} & \textbf{0.98} & \textbf{0.98} \\
         & Claude & \textbf{0.97} & \textbf{0.99} & \textbf{0.95} & \textbf{0.97} \\
        \hline
    \end{tabular}
\end{table}

In this section, we compare our method (CEF-Log) with standard and state-of-the-art (SoTA) prompting methods on the CSIC 2010 dataset using Gemini Flash 2.5. Additionally, we conduct cross-validation with Claude 3.7 Sonet to confirm the effectiveness of our proposed method. From Tables \ref{tab:standard_prompting_baselines}-\ref{tab:sota_prompting_comparison}, we observe the following key findings::
\begin{itemize}
    \item \textbf{CEF-Log achieves superior performance with minimal examples.} As shown in Table \ref{tab:standard_prompting_baselines}, standard few-shot prompting requires $k=40$ examples to reach $96\%$ recall and $98\%$ F1-score. In contrast, CEF-Log achieves comparable performance using only four examples, representing a $10\times$ improvement in sample efficiency through structured reasoning methodology rather than pattern memorization. 
    \item \textbf{Cross-model validation confirms method generalizability.} Table \ref{tab:cross_validation} demonstrates that CEF-Log's improvements transfer across LLM architectures. The refined Chain-of-Thought baseline without domain-specific context shows substantial divergence between models (F1=0.95 for Gemini vs. 0.78 for Claude), indicating that generic reasoning prompts are insufficient without forensic domain knowledge integration.
    \item \textbf{CEF-Log outperforms existing prompting-based methods.} Table \ref{tab:sota_prompting_comparison} shows that, compared to LogGPT \cite{qi2023loggpt} and LogPrompt \cite{liu2024logprompt}, CEF-Log can achieve perfect performance while requiring only 4 log examples.
\end{itemize}

\begin{table}[h]
\centering
\caption{Performance Comparison of State-of-the-art Prompting Methods}
\label{tab:sota_prompting_comparison}
\begin{tabular}{lcccc}
\hline
\textbf{Method} & \textbf{Samples} & \textbf{Prec.} & \textbf{Rec.} & \textbf{F1} \\
\hline
LogGPT & 10 & 0.587 & 1.0 & 0.74 \\
LogPrompt & 20 & 0.31 & 0.87 & 0.65 \\
CEF-Log (Ours) & 4 & 0.99 & 0.98 & 0.98 \\
\hline
\end{tabular}
\end{table}

\subsubsection{Performance Comparison with ML Baselines} 
Table~\ref{tab:ml_comparison} compares CEF-Log against traditional ML classifiers trained on 40 labeled examples per seed. All ML baselines exhibit substantially lower performance: Random Forest achieves the highest ML accuracy at 67.2\%, followed by Logistic Regression (63.8\%) and SVM (62.4\%). The performance gap is particularly pronounced in F1-score, where CEF-Log (0.97) outperforms Random Forest (0.67) by 30 percentage points.

This disparity stems from the fundamental limitation of ML approaches on small training sets: 40 examples provide insufficient coverage of the diverse attack patterns in web server logs. In contrast, CEF-Log leverages the LLM's pre-trained knowledge of attack semantics, using few-shot examples to demonstrate analytical methodology rather than exhaustively enumerate attack signatures.

\begin{table}[ht]
    \centering
    \caption{Comparison with Machine Learning Baselines (Gemini Flash 2.5, 40 Training Examples)}
    \label{tab:ml_comparison}
    \begin{tabular}{lcccc}
        \hline
        \textbf{Method} & \textbf{Acc.} & \textbf{Prec.} & \textbf{Rec.} & \textbf{F1} \\
        \hline
        SVM & 0.624 & 0.648 & 0.624 & 0.595 \\
        Random Forest & 0.672 & 0.683 & 0.672 & 0.667 \\
        Logistic Regression & 0.638 & 0.646 & 0.638 & 0.633 \\
        \textbf{CEF-Log ($k$=4)} & \textbf{0.97} & \textbf{0.99} & \textbf{0.95} & \textbf{0.97} \\
        \hline
    \end{tabular}
\end{table}

\subsubsection{Sample Efficiency Comparison with Few-shot Prompting Baseline}

\begin{figure}[ht]
    \centering
    \includegraphics[width=\columnwidth]{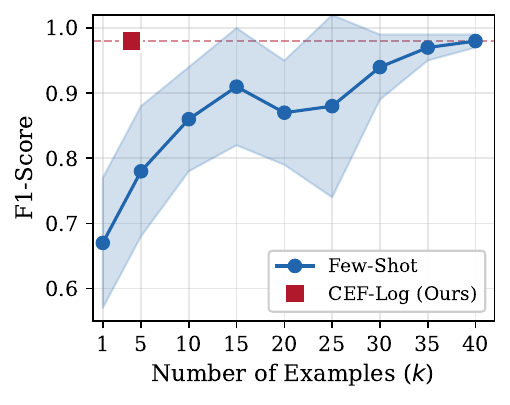}
    \caption{Sample efficiency comparison between standard few-shot prompting and CEF-Log. Few-shot requires $k$=40 examples to achieve F1=0.98, while CEF-Log matches this performance with only $k$=4 examples. Shaded regions indicate standard deviation across five seeds.}
    \label{fig:sample_efficiency}
\end{figure}

A critical advantage of CEF-Log is sample efficiency while achieving high performance with minimal labeled data. Fig.~\ref{fig:sample_efficiency} illustrates the relationship between example count and detection performance for standard few-shot prompting versus CEF-Log.

Few-shot prompting exhibits monotonic improvement with example count: F1-score increases from 0.67 ($k$=1) to 0.98 ($k$=40), with diminishing returns beyond $k$=30. CEF-Log achieves comparable performance (F1=0.98) with only four examples, which is a 10$\times$ reduction in labeling requirements. This efficiency gain is attributed to the reasoning template, which teaches the LLM \textit{how} to analyze logs rather than \textit{what} patterns to memorize. 

\subsubsection{Performance on ForenWebLog Dataset} 
We evaluate CEF-Log on the ForenWebLog dataset using only four examples. Table~\ref{tab:realworld} presents results across five sampling seeds.

CEF-Log achieves 96\% mean accuracy with perfect recall (100\%), correctly identifying all attack instances. The 4\% accuracy gap stems from false positives on benign traffic patterns not represented in CSIC 2010 examples. Notably, the model correctly classifies all false positive candidates (HTTPS-on-HTTP attempts), demonstrating robust discrimination between suspicious-appearing benign traffic and actual attacks.

\begin{table}[t]
\centering
\caption{CEF-Log Performance on ForenWebLog Dataset}
\label{tab:realworld}
\begin{tabular}{lcccc}
\hline
\textbf{Seed} & \textbf{Acc.} & \textbf{Prec.} & \textbf{Rec.} & \textbf{F1} \\
\hline
Seed 1 & 1.00 & 1.00 & 1.00 & 1.00 \\
Seed 2 & 0.96 & 0.93 & 1.00 & 0.96 \\
Seed 3 & 0.95 & 0.90 & 1.00 & 0.95 \\
Seed 4 & 0.95 & 0.90 & 1.00 & 0.95 \\
Seed 5 & 0.95 & 0.90 & 1.00 & 0.95 \\
\hline
\textbf{Mean $\pm$ Std} & 0.96 $\pm$ 0.02 & 0.93 $\pm$ 0.04 & 1.00 $\pm$ 0.00 & 0.96 $\pm$ 0.02 \\
\hline
\end{tabular}
\end{table}

\subsection{Analysis}
This section provides an analysis of CEF-Log's performance characteristics, examining temperature sensitivity, reasoning quality, and multi-step attack detection capabilities.

\subsubsection{Analysis on Temperature Tuning}
We evaluated the effect of the temperature parameter $\tau$ on classification performance, varying from 0.2 to 2.0 across five random seeds per configuration (Table \ref{tab:temperature}). Results show that accuracy remains stable ($97-99\%$) across all temperature values, with precision consistently near 100\%. However, higher temperatures correlate with increased output variability, as evidenced by standard deviation in recall increasing from 0.05 ($\tau=1.0$) to 0.14 ($\tau=2.0$). We adopt $\tau = 1.0$ for all experiments, as it provides the lowest variance while maintaining competitive performance.

\begin{table}[htbp]
    \centering
    \caption{Effect of Temperature on Classification Performance on ForenWebLog Dataset}
    \label{tab:temperature}
    \begin{tabular}{ccccc}
        \hline
        \textbf{Temp.} & \textbf{Acc.} & \textbf{Prec.} & \textbf{Rec.} & \textbf{F1} \\
        \hline
        0.2 & 0.97 $\pm$ 0.05 & 1.00 $\pm$ 0.00 & 0.95 $\pm$ 0.09 & 0.96 $\pm$ 0.08 \\
        0.6 & 0.97 $\pm$ 0.04 & 1.00 $\pm$ 0.00 & 0.95 $\pm$ 0.07 & 0.97 $\pm$ 0.06 \\
        1.0 & 0.96 $\pm$ 0.02 & 1.00 $\pm$ 0.00 & 0.93 $\pm$ 0.05 & 0.95 $\pm$ 0.04 \\
        1.4 & 0.99 $\pm$ 0.06 & 1.00 $\pm$ 0.00 & 0.96 $\pm$ 0.13 & 0.96 $\pm$ 0.10 \\
        2.0 & 0.99 $\pm$ 0.06 & 1.00 $\pm$ 0.00 & 0.98 $\pm$ 0.12 & 0.98 $\pm$ 0.10 \\
        \hline
    \end{tabular}
\end{table}

\subsubsection{Analysis on LLM-Reasoning}
A distinguishing feature of LLM-based forensic analysis is the generation of human-readable explanations accompanying each classification. To evaluate reasoning quality, we conducted a qualitative assessment using 20 manually selected log entries (10 benign, 10 malicious) from the ForenWebLog dataset, representing diverse attack categories including Log4Shell exploitation, SQL injection, XSS, command injection, and web shell deployment.

A forensic investigation expert evaluated each explanation using a 5-point scale: (1) incorrect or absent reasoning, (2) partially correct but leading to wrong classification, (3) partially correct with correct classification, (4) correct with adequate detail, and (5) correct, detailed, and forensically sound. Table~\ref{tab:reasoning_quality} summarizes the reasoning quality assessment. Both models demonstrated strong reasoning capabilities, with average scores of 4.65 (Gemini) and 4.55 (Claude) out of 5.0. Notably, both models correctly identified sophisticated attack patterns, including:

\begin{itemize}
    \item \textbf{Log4Shell (CVE-2021-44228)}: Both models recognized JNDI lookup patterns (\texttt{\$\{jndi:ldap://...\}}) across multiple HTTP fields and correctly attributed the attack to Log4j exploitation.
    \item \textbf{Shellshock (CVE-2014-6271)}: Models identified bash vulnerability exploitation in User-Agent headers with command injection payloads.
    \item \textbf{Obfuscated XSS}: Despite heavy URL encoding, both models decoded and identified JavaScript injection attempts.
    \item \textbf{Time-based SQL Injection}: Models correctly identified \texttt{sleep()} functions as blind SQL injection techniques.
\end{itemize}

\begin{table}[htbp]
\centering
\caption{Reasoning Quality Assessment (Mean Score $\pm$ Std)}
\label{tab:reasoning_quality}
\begin{tabular}{lcc}
\hline
\textbf{Attack Category} & \textbf{Gemini} & \textbf{Claude} \\
\hline
Injection Attacks (SQL, XSS, Cmd) & 4.80 $\pm$ 0.45 & 4.40 $\pm$ 0.89 \\
Known CVEs (Log4Shell, Shellshock) & 5.00 $\pm$ 0.00 & 5.00 $\pm$ 0.00 \\
Web Shell Sequences & 4.00 $\pm$ 0.82 & 4.50 $\pm$ 0.58 \\
Benign Traffic & 5.00 $\pm$ 0.00 & 4.70 $\pm$ 0.48 \\
\hline
\textbf{Overall} & \textbf{4.65 $\pm$ 0.49} & \textbf{4.55 $\pm$ 0.60} \\
\hline
\end{tabular}
\end{table}

The explanations generated by CEF-Log satisfy key forensic requirements: they reference specific elements within the log entry (traceability), correctly identify attack vectors when present (accuracy), and provide reasoning that could be presented in legal proceedings (admissibility). This represents a significant advantage over traditional ML approaches, which provide only binary labels without interpretable justification.

\subsubsection{Analysis on Multi-step Attack Patterns Recognition}
We designed experiments to assess CEF-Log's contextual reasoning capabilities across three attack categories: Denial of Service (DoS), brute force authentication attacks, and web shell deployment sequences. For each attack category, we created three scenario variants with attack requests interspersed among 50 benign entries. 

Both Gemini Flash 2.5 and Claude 3.7 Sonnet correctly classified all attack instances across all scenarios (100\% recall). More significantly, the reasoning traces demonstrated genuine contextual understanding rather than isolated pattern matching.

\textbf{DoS Detection}: Both models identified the coordinated nature of flooding attacks. Gemini's reasoning for DOS01 stated: ``\textit{Repeated POST requests to /xmlrpc.php from the same IP, indicative of a brute-force or denial-of-service attempt against a WordPress site.}'' Claude similarly noted the pattern continuation: ``\textit{Part of the ongoing flood attack pattern from IP 167.138.163.110 targeting /impressum.}''

\textbf{Brute Force Detection}: Models correctly identified authentication attack patterns by correlating multiple login attempts from identical source addresses. For BRUTEFORCE03, Claude's reasoning explicitly referenced the attack sequence: ``\textit{Continued login attempts from IP 225.65.201.243 with very short time interval from previous attempt.}''

\textbf{Web Shell Detection}: This category tested the most sophisticated contextual reasoning, requiring correlation between an initial exploitation phase and subsequent payload execution. Results were mixed regarding cross-reference quality:

\begin{itemize}
    \item \textbf{SHELL01} (path traversal + execution): Claude correctly linked the execution attempt to the prior upload: ``\textit{Follow-up attempt to the attack in Record ID 66, trying to access a potentially created malicious file in the temp directory.}''
    \item \textbf{SHELL02} (WebDAV + execution): Both models classified correctly but did not explicitly reference the relationship between upload and execution phases.
    \item \textbf{SHELL03} (Gravity Forms exploit + execution): Gemini demonstrated superior cross-referencing: ``\textit{GET request for '/uploads/picture.jpg.php' [...] is likely an attempt to activate a web shell, likely an attempt to activate a web shell.}''
\end{itemize}

Table~\ref{tab:multistep_results} summarizes the multi-step attack detection results.

\begin{table}[htbp]
\centering
\caption{Multi-step Attack Detection Performance}
\label{tab:multistep_results}
\begin{tabular}{lccc}
\hline
\textbf{Scenario} & \textbf{Detection} & \textbf{Context} & \textbf{Cross-Ref} \\
\textbf{Type} & \textbf{Rate} & \textbf{Aware} & \textbf{Quality} \\
\hline
DoS (3 variants) & 100\% & Yes & High \\
Brute Force (3 variants) & 100\% & Yes & High \\
Web Shell (3 variants) & 100\% & Partial & Medium \\
\hline
\end{tabular}
\end{table}

\section{Conclusion} \label{sec:conclusion}
This paper presented CEF-Log, a context-enhanced few-shot chain-of-thought prompting strategy that enables LLMs to perform forensic web server log analysis with high accuracy and explainable reasoning. By encoding expert investigative methodology through a structured five-step reasoning template, CEF-Log teaches the model how to analyze logs rather than what patterns to memorize. Experimental results demonstrate that CEF-Log achieves an F1-score of 0.99 on CSIC 2010 using only four examples, a $10\times$ improvement in sample efficiency over standard few-shot prompting. Furthermore, our method also substantially outperform traditional ML baselines. Cross-validation across Gemini Flash 2.5 and Claude 3.7 Sonnet confirms that gains stem from the methodology rather than model-specific characteristics. We also introduced ForenWebLog, a realistic dataset with multi-step attack sequences, on which CEF-Log achieved perfect recall. Qualitative evaluation confirmed that generated explanations satisfy forensic traceability and accuracy requirements for legal proceedings.

Future work will address current limitations, including broader dataset validation, improved cross-referencing in multi-step attack detection, and reduced inference costs through model distillation. We also plan to integrate retrieval-augmented generation for dynamic threat intelligence incorporation and extend the methodology to additional forensic domains.

% \section*{Acknowledgments}
% This should be a simple paragraph before the References.

% --- BIBLIOGRAPHY SECTION ---
% 1. Set the style to IEEEtran
\bibliographystyle{IEEEtran}
% 2. Specify the name of your .bib file (without the extension)
\bibliography{references} 

% \newpage

% \section{Biography Section}
% \begin{IEEEbiographynophoto}{John Doe}
% Biography text here.
% \end{IEEEbiographynophoto}

\vfill
\end{document}